\documentclass[12pt]{article}
\usepackage{amssymb}

\newcommand{\be}{\begin{equation}}
\newcommand{\ee}{\end{equation}}

\newcommand{\oj}{{\overline j}}

\newcommand{\p}{\partial}
\newcommand{\ta}{\tilde{A}}
\newcommand{\tb}{\tilde{B}}

\newcommand{\nn}{\nonumber}
\newcommand{\emn}{\epsilon_{\mu\nu\gamma}}

\newcommand{\no}{\noindent}

\def\bea{\begin{eqnarray}}
\def\eea{\end{eqnarray}}
\def\beq{\begin{eqnarray}}
\def\eeq{\end{eqnarray}}

\begin{document}

\title{\textbf{Generalized duality between local vector theories in $D=2+1$}}
\author{D. Dalmazi \\
\textit{{UNESP - Campus de Guaratinguet\'a - DFQ} }\\
\textit{{Av. Dr. Ariberto Pereira da Cunha, 333} }\\
\textit{{CEP 12516-410 - Guaratinguet\'a - SP - Brazil.} }\\
\textsf{E-mail: dalmazi@feg.unesp.br }}
\date{\today}
\maketitle

\begin{abstract}

The existence of an interpolating  master action does not guarantee the same spectrum
for the interpolated dual theories. In the specific case of a generalized self-dual
(GSD) model defined as the addition of the Maxwell term to the self-dual model in
$D=2+1$, previous master actions have furnished a dual gauge theory which is either
nonlocal or contains a ghost mode. Here we show that by reducing the Maxwell term to
first order by means of an auxiliary field we are able to define a master action which
interpolates between the GSD model and a couple of non-interacting
Maxwell-Chern-Simons theories of opposite helicities. The presence of an auxiliary
field explains the doubling of fields in the dual gauge theory. A generalized duality
transformation is defined and both models can be interpreted as self-dual models.
Furthermore, it is shown how to obtain the gauge invariant correlators of the
non-interacting MCS theories from the correlators of the self-dual field in the GSD
model and vice-versa. The derivation of the non-interacting MCS theories from the GSD
model, as presented here, works in the opposite direction of the soldering approach.
 \textit{{PACS-No.:}
11.15.-q, 11.10.Kk, 11.10.Gh, 11.10.Ef }
\end{abstract}

\newpage

\section{Introduction}

The existence of different but equivalent descriptions of the same physical theory can
help us to reveal deep aspects of the theory which are apparent in one formulation but
hidden in the other one. One successful example is the bosonization program in $1+1$
dimensions \cite{Coleman,Abdalla}. Recent examples are provided by the AdS/CFT
correspondence \cite{JM} and the work of \cite{SW} where duality played a key role in
a rigorous proof of confinement in a four dimensional theory. A simple approach for
deriving dual theories at quantum level is the use of interpolating master actions
\cite{DJ}, see \cite{HL} for a review. In \cite{DJ} a first order master action was
suggested in order to prove duality equivalence between a non-gauge theory of the
self-dual (SD) type \cite{TPN} (first order) and a second order Maxwell-Chern-Simons
(MCS) theory. Both theories represent  one massive polarization state in $D=2+1$
spacetime  of helicity $+1$ or $-1$, depending on the sign of the Chern-Simons
coefficient. As expected from the lack of gauge invariance of the SD theory, the map
between the theories $f_{\mu}\leftrightarrow \frac{\emn \p^{\nu}A^{\gamma}}m $ is
invariant under gauge transformations of the Maxwell-Chern-Simons fundamental field
$A^{\gamma}$ and holds at classical and quantum level including gauge invariant
correlation functions \cite{BRR}. The natural addition of a Maxwell term in the
self-dual model however, spoils the simplicity of the duality relation between the now
called generalized self-dual (GSD) model and its possible gauge invariant dual theory.
A direct generalization of the master action approach leads, quite surprisingly, to a
gauge theory \cite{Bazeiajpa} which now includes a ghost mode in the spectrum, so the
existence of a master action which interpolates between two theories does not
guarantee spectrum equivalence {\it a priori}. As explained in \cite{jhep} if we
insist in the spectrum equivalence a new master action can be suggested which leads
however, to a non-local vector theory. It is seems that we have glanced the old
problem of formulating massive theories in a gauge invariant way. In this work we show
that by introducing an auxiliary vector field to lower the Maxwell term to first order
we are able to define another master action which naturally interpolates between the
GSD model and a well defined gauge invariant local theory which corresponds to a
couple of non-interacting Maxwell-Chern-Simons theories of opposite helicities,
henceforth called 2MCS. It turns out that the GSD model and the 2MCS models were know
to be related for a long time \cite{DJT,Deser}. In particular, it has been shown in
\cite{bw527,BK1} that the two MCS models could be soldered into the GSD model. Our
results are complementary to the soldering procedure and work in the opposite
direction like the canonical transformations of  \cite{BG,BK2}. In the next section we
quickly review previous master action attempts and suggest a new master action and a
generating functional which allows us to compare correlation functions in both
theories. In section III we concentrate on the classical equivalence, clarifying how a
theory of two non-interacting vector fields can be shown to be physically equivalent
to the GSD model which contains only one vector field. We also comment in section III
on the coupling to matter fields. In section IV we draw our conclusions.

\section{Master action and quantum equivalence}

Let us first present the GSD model which might be called also a
Maxwell-Chern-Simons-Proca model\footnote{Comparing to \cite{jhep} we have slightly
changed our notation  $a_2 \to - a_2$ but we still use $g_{\mu\nu}=(+,-,-)$.} :

\be {\cal L}_{GSD} \, = \, a_0 f^{\mu}f_{\mu} + a_1
\epsilon_{\alpha\beta\gamma}f^{\alpha}\partial^{\beta}f^{\gamma} - \frac{a_2}2
F_{\mu\nu}(f)F^{\mu\nu}(f) \label{gsd} \ee

\no For $a_2=0$ we recover the self-dual model of \cite{TPN}. Due to unitarity reasons
we need to have \cite{Oswaldo} ( see also \cite{jhep}) $a_0 \ge 0 $ and $a_2 \ge 0$.
The constants $a_i$ are otherwise arbitrary. Henceforth we assume that $a_0$ and $a_2$
are definite positive. We can write down the equations of motion of (\ref{gsd}) in a
self-dual form generalizing the definition of duality transformations:

\be f_{\mu} = \frac 1{a_0}\left( a_1 E_{\mu\nu} - a_2 \Box\theta_{\mu\nu} \right)
f^{\nu} \equiv f^*. \label{eqm} \ee

\no We have defined the differential operators:

\be E_{\mu\nu} = \emn \p^{\gamma} \quad , \quad \Box\theta_{\mu\nu} = \Box g_{\mu\nu}
- \p_{\mu}\p_{\nu}\label{et}\ee

\no Note the useful identities $E_{\mu\nu}E^{\nu\alpha}=- \Box \theta_{\mu}^{\alpha}$,
$ E_{\mu\nu}\theta^{\nu\alpha}= E _{\mu}^{\alpha}$ and
$\theta_{\alpha\beta}\theta^{\beta\gamma}=\theta_{\alpha}^{\gamma}$. From (\ref{eqm})
we can derive the existence of two massive modes in the self-dual field: $\left( \Box
+ m_+^2 \right) \left(\Box + m_-^2 \right)f_{\mu} =0$, where

\be 2 m_{\pm}^2 \, =\, b^2 + 2 a \pm \sqrt{(b^2 + 2a)^2-4 a^2} \quad , \label{mpm} \ee

\no with  $a=a_0/a_2 \, , \, b=a_1/a_2$. Those massive physical particles can be
confirmed by checking the poles and the corresponding signs of the residues of the
propagator. The expression (\ref{mpm}) can be inverted for the ratios $a,b$:

\be a_0 = a_2 \, m_+ m_- \quad ; \quad a_1 = a_2 \, ( m_+ - m_-) \quad . \label{ab}\ee

\no There is a sign freedom in the solution for $a_1$ but we choose it to be positive
for definiteness. Henceforth we can describe the GSD model as defined by the three
parameters $a_2,m_+,m_-$ according with (\ref{gsd}) and (\ref{ab}). This is a more
physical notation which makes clear, in particular, that the mass split comes from
parity breaking. If $a_1=0$ we have the Maxwell-Proca theory with two particles with
opposite helicities $\pm 1$ but with degenerate mass $m_+=m_-$.

In order to suggest a new master action which would produce a local gauge theory dual
to (\ref{gsd}) we recall previous attempts. Both suggestions of \cite{Bazeiajpa} and
\cite{jhep} can be cast in the form of a gauge invariant second order master equation:

\beq {\cal L} \, &=& \, a_0 f^{\mu}f_{\mu} + a_1
\epsilon_{\alpha\beta\gamma}f^{\alpha}\partial^{\beta}f^{\gamma} - \frac{a_2}2
F_{\mu\nu}(f)F^{\mu\nu}(f) \nn\\
&-& b_1 \epsilon_{\alpha\beta\gamma}(A^{\alpha}-
f^{\alpha})\partial^{\beta}(A^{\gamma}-f^{\gamma}) + \frac{b_2}2
F_{\mu\nu}(A-f)F^{\mu\nu}(A-f) \label{master1} \eeq

\no The proposal of \cite{Bazeiajpa} corresponds to $(b_1,b_2)=(a_1,a_2)$. The
advantage of this choice is that all quadratic terms in the self-dual field except the
first one on the right-handed side of (\ref{master1}) are cancelled, which gives rise
to a local gauge theory upon integration in the self-dual field. However, the theory
thus obtained contains a ghost pole in the propagator, which is in agreement with the
predictions of \cite{Baeta1}. The presence of the ghost could have been foreseen also
from the fact that after a trivial shift $A_{\mu} \to A_{\mu} + f_{\mu}$, the theory
(\ref{master1}) can be written as a GSD model decoupled from a Maxwell-Chern-Simons
Gauge theory where the coefficient of the Maxwell term appears with the wrong sign.
The integration on the self-dual field reintroduces, as explained in \cite{jhep}, the
ghost mode in the resulting gauge theory. So the message is clear, i.e., we better mix
the self-dual and the gauge field through a Lagrangian density which has no particle
content thus guaranteeing the spectrum match of both gauge and non-gauge theories.
This the case of the choice $(b_1,b_2)=(a_1,0)$ where the mixing comes only from the
topological Chern-Simons term which contains no physical degree of freedom. Indeed,
this choice leads to a gauge theory \cite{jhep} equivalent to the GSD model, up to
contact terms in the correlation functions, and with the same massive poles
$k^2=m_{\pm}^2$ without extra particles in the spectrum. Due to the non-cancelation of
the quadratic terms in the self-dual field which involve derivatives, we pay the price
of loosing locality upon integration on the self-dual field. A key ingredient lacking
in (\ref{master1}) but present in the original proposal of a master action in
\cite{DJ} is to start with a first order Lagrangian. It is easy to reduce the Maxwell
term to first order by using an auxiliary vector field ($g_{\mu}$), such that we are
led to the following suggestion:

\beq {\cal L}_{Master} \, &=& \, a_0 f^{\mu}f_{\mu} + a_1
\epsilon_{\alpha\beta\gamma}f^{\alpha}\partial^{\beta}f^{\gamma} + g^{\mu} g_{\mu} +
\sqrt{a_2} \, g_{\mu} \epsilon^{\mu\alpha\beta}  F_{\alpha\beta}(f) + f_{\mu} j^{\mu}\nn\\
&-& a_1 \epsilon_{\alpha\beta\gamma}(\ta^{\alpha}- f^{\alpha})\p^{\beta}(\ta^{\gamma}-
f^{\gamma}) - \sqrt{a_2}\, (\tb_{\mu} - g_{\mu})\epsilon^{\mu\alpha\beta}
F_{\alpha\beta}(\ta- f)
 \label{master2} \eeq

\no For $a_2=0$ the auxiliary field $g_{\mu}$ decouples and we recover the master
action of \cite{DJ} plus a source term for the self-dual field that we have introduced
for future use. After the shifts $\ta_{\mu} \to \ta_{\mu} + f_{\mu}$ and $\tb_{\mu}
\to \tb_{\mu} + g_{\mu}$ we end up with the GSD model decoupled from a trivial
topological theory for the fields $\ta_{\mu}$ and $\tb_{\mu}$ with no particle
content. If we finally integrate over $\ta_{\mu},\tb_{\mu}$ and $g_{\mu}$ in the path
integral we derive the GSD model (\ref{gsd}) plus a source term:

\be {\cal Z}(j) = \int  {\cal D}f^{\nu} {\cal D}g^{\nu} {\cal D}\ta^{\nu} {\cal
D}\tb^{\nu} \, e^{ i\int d^3x {\cal L}_{{\rm Master}}} \, = \, \int  {\cal D}f^{\nu}
\, e^{i\int d^3x\, \left({\cal L}_{GSD} + j_{\mu} f^{\mu}\right)} \label{zj1} \ee

\no On the other hand, since in the master action (\ref{master2}) there are no
quadratic terms in the fields $f_{\mu}$ and $g_{\mu}$ involving derivatives , they can
be easily integrated over such that we are left with a local gauge theory
corresponding to a couple of interacting Maxwell-Chern-Simons models plus source
dependent terms:

\bea {\cal Z}(j) &=& \int  {\cal D}f^{\nu} {\cal D}g^{\nu} {\cal D}\ta^{\nu} {\cal
D}\tb^{\nu} \, \exp i\int d^3x {\cal L}_{{\rm
Master}}\nn \\
&=& \int  {\cal D}\ta^{\nu}{\cal D}\tb^{\nu}  \, \exp \, i\int d^3x\, \left\lbrack
\tilde{{\cal L}}(\ta,\tb) - \frac {j_{\mu} j^{\mu}}{4 a_0}
 - \frac {j^{\mu}\emn F^{\nu\gamma}(
a_1 \ta + \sqrt{a_2} \tb )}{2 a_0}  \right\rbrack \label{zj2} \eea

\no where

\be \tilde{{\cal L}}(\ta,\tb) = - \frac 1{2 a_0} F_{\alpha\beta}^2 (a_1 \ta +
\sqrt{a_2} \tb ) - \frac{a_2}2 F_{\alpha\beta}^2 (\ta ) - a_1 \emn \ta^{\mu}
\p^{\nu}\ta^{\gamma} - 2 \sqrt{a_2} \emn \ta^{\mu} \p^{\nu}\tb^{\gamma} \label{ltilde}
\ee

\no After appropriate field redefinitions we can rewrite $\tilde{{\cal L}}(\ta,\tb)$
as a couple of non-interacting Maxwell-Chern-Simons theories. For instance, using

\bea \ta_{\mu} &=& \frac 1{\sqrt{2 a_2 (m_+ + m_-)}}\left(
\sqrt{m_+} A_{\mu} - \sqrt{m_-} B_{\mu} \right) \nn \\
\label{linear}\\
 \tb_{\mu} &=& \frac {-1}{\sqrt{2 (m_+ +
m_-)}}\left( \sqrt{m_+^3} A_{\mu} + \sqrt{m_-^3} B_{\mu} \right) \nn\eea

\no We have

\be \tilde{{\cal L}}(\ta,\tb) = {\cal L}_{2MCS}(A,B) = -\frac{F_{\alpha\beta}^2 (A)}4
+ \frac {m_+}2 \emn A^{\mu}\p^{\nu}A^{\gamma} -\frac{F_{\alpha\beta}^2 (B)}4 - \frac
{m_-}2 \emn B^{\mu}\p^{\nu}B^{\gamma} \label{2mcs} \ee

\no The field redefinitions (\ref{linear}) are not unique but the other possible
choices also lead to the same non-interacting Chern-Simons theories (\ref{2mcs}) up to
trivial field rescalings. In terms of the new fields we can rewrite (\ref{zj2}), up to
a trivial constant Jacobian, as follows:

\be {\cal Z}(j) = \int  {\cal D}A^{\nu}{\cal D}B^{\nu}  \, \exp \left\lbrace i\int
d^3x\, \left\lbrack  {\cal L}_{2MCS}(A,B) - \frac { j_{\mu} j^{\mu}}{4 a_2\, m_+ m_-}
+ j^{\mu} C_{\mu} \right\rbrack \right\rbrace \label{zj3} \ee

\no Where we have defined the gauge invariant combination

\be C_{\mu} = - \frac 1{2 a_0}\emn F^{\nu\gamma}(a_1 \ta + \sqrt{a_2}\tb ) =
\frac{\emn\p^{\nu}}{\sqrt{a_2(m_+ + m_-)}}\left(\frac{A^{\gamma}}{\sqrt{m_+}} +
\frac{B^{\gamma}}{\sqrt{m_-}}\right) \label{c}\ee

\no Deriving (\ref{zj3}) and  (\ref{zj1}) with respect to the sources we have the
equivalence of correlation functions:

\be \left\langle f_{\mu_1}(x_1) \cdots f_{\mu_N}(x_N) \right\rangle_{GSD} =
\left\langle C_{\mu_1}(x_1) \cdots C_{\mu_N}(x_N) \right\rangle_{2MCS} \, + \, {\rm
contact} \, \, {\rm terms} \quad . \label{cf}\ee

\no Where the contact terms (delta functions) come from the quadratic term in the
sources appearing in (\ref{zj3}). As expected from the fact that the GSD model is not
a gauge theory, we have identified correlation functions of the self-dual field with
correlation functions of a gauge invariant object in the 2MCS theory with no need of
introducing an explicit gauge condition. By examining the propagators of the fields
$A_{\mu}$ and  $B_{\mu}$ in the 2MCS model we notice that that both have a pole at
momenta $k^2=0$ which represents in fact a non-propagating mode (vanishing residue
\cite{Baeta1}) and a physical pole at $k^2=m_+^2$ and $k^2=m_-^2$ respectively.
Therefore the spectrum of the 2MCS and the GSD models are equivalent as expected.
However, it is rather disturbing for a complete proof of equivalence of such models
that the correlation functions of the self-dual field can be written in terms of
correlation functions of only one specific linear combination of $A_{\mu}$ and
$B_{\mu}$ fields which are on their turn independent and non-interacting fields and
can not be written of course in terms of just one linear combination. It is natural to
ask whether correlation functions of both fields $A_{\mu},B_{\mu}$ can be in general
calculated from the GSD theory. In order to answer that question we define a new
generating function below which allows the computation of  the relevant gauge
invariant correlators of the 2MCS theory:

\bea {\cal Z}(j_A,j_B) &=& \int  {\cal D}f^{\nu} {\cal D}g^{\nu} {\cal D}\ta^{\nu}
{\cal D}\tb^{\nu} \, \exp\left\lbrace i\int d^3x \left\lbrack {\cal L}_{{\rm
Master}} + j^{\mu}_A F_{\mu}(A) + j^{\mu}_B F_{\mu}(B)\right\rbrack \right\rbrace \label{zjab} \\
&=& \int  {\cal D}f^{\nu} {\cal D}g^{\nu} {\cal D}\ta^{\nu} {\cal D}\tb^{\nu} \,
\exp\left\lbrace  i\int d^3x \left\lbrack {\cal L}_{{\rm Master}} + r\left( m_-
\oj_{A}^{\mu} - m_+\oj_{B}^{\mu}\right) F_{\mu}(\ta )\right.\right. \nn\\ &-& \left.
\left. \frac r{\sqrt{a_2}} \left(\oj_{A}^{\mu} + \oj_{B}^{\mu}\right) F_{\mu}(\tb
)\right\rbrack \right\rbrace \label{zjatbt} \eea

\no Where we have introduced the constant $\, r= \sqrt{2a_2/(m_+ + m_-)} \,$, the dual
field strength $F_{\mu}(A)=\emn\p^{\nu}A^{\gamma}$ and the redefined sources
$\oj^{\mu}_A = j^{\mu}_A/\sqrt{m_+} \, ; \, \oj^{\mu}_B = j^{\mu}_B/\sqrt{m_-}$. In
obtaining (\ref{zjatbt}) from (\ref{zjab}) we have inverted the linear transformations
(\ref{linear}). Since the 2MCS model follows from the master action by integrating
over ${\cal D}f^{\nu} {\cal D}g^{\nu}$ it is clear that $j^{\mu}_A $ and $j^{\nu}_B$
are the correct sources for computing correlation functions of $F^{\mu}(A)$ and
$F^{\nu}(B)$ in the 2MCS theory respectively. Now if we integrate over ${\cal
D}g^{\nu} {\cal D}\ta^{\nu} {\cal D}\tb^{\nu}$ we deduce:

\bea {\cal Z}(j_A,j_B) &=& \int  {\cal D}f^{\nu} \, \exp i\int d^3x \left\lbrack {\cal
L}_{GSD} - r \left(\oj^{\mu}_A + \oj^{\mu}_B\right)\Box\theta_{\mu\nu}f^{\nu}- \frac
1{4(m_+ + m_-)}F_{\alpha\beta}^2\left(\oj^{\mu}_A +
\oj^{\mu}_B\right)\right.\nn\\
&+& \left. r \left(m_- \oj^{\mu}_A - m_+ \oj^{\mu}_B\right) \emn\p^{\nu} f^{\gamma} -
\frac 12 \left(\oj^{\mu}_A - \oj^{\mu}_B\right)\emn\p^{\nu} \left(\oj^{\gamma}_A +
\oj^{\gamma}_B\right)\right\rbrack  \label{zj4} \eea

\no In conclusion, we can indeed calculate correlation functions of  the 2MCS theory
from the GSD model. Explicitly,

 \bea  &
&\left\langle F_{\mu_1}\lbrack A(x_1)\rbrack \cdots F_{\mu_N}\lbrack A(x_N)\rbrack
F_{\nu_1}\lbrack B(y_1)\rbrack \cdots F_{\nu_N}\lbrack B(y_M)\rbrack
\right\rangle_{2MCS}
\nn\\
&=& \hat{T}_{\mu_1\alpha_1}(m_-,x_1) \cdots
\hat{T}_{\mu_N\alpha_N}(m_-,x_N)\hat{T}_{\nu_1\beta_1}(-m_+,y_1)\cdots
\hat{T}_{\nu_M\beta_M}(-m_+,y_M) \times \nn \\
& & \times \left\langle f^{\alpha_1}(x_1)\cdots f^{\alpha_N}(x_N)
f^{\beta_1}(y_1)\cdots f^{\beta_M}(y_M) \right\rangle_{GSD} \, + \, {\rm contact} \,
{\rm terms} \label{cf2} \eea

\no Where $\hat{T}_{\alpha\beta}(m,x)= -\left( r/\sqrt{\vert
m\vert}\right)\left(\Box\theta_{\alpha\beta} + m E_{\alpha\beta}
\right)_x$. One can check, as we have done, the correctness of
(\ref{zj4}) and (\ref{cf2}) by calculating two point functions in
the 2MCS theory from the self-dual propagator in the GSD theory
plus the contact terms. In particular, the contact terms are such
that one verifies the trivial result $\left\langle F_{\mu}\lbrack
A(x)\rbrack F_{\nu}\lbrack B(y)\rbrack \right\rangle_{2MCS} =0 $.

The results (\ref{cf}) and (\ref{cf2}) demonstrate the quantum equivalence of the
gauge invariant sector of the 2MCS model to the GSD model. The equivalence holds up to
contact terms which vanish for non-coinciding points.

\section{Classical equivalence and generalized self-duality}

From the master action ${\cal L}_{Master}(f,g,\ta,\tb)$ given in (\ref{master2}) we
have the following equations of motion:

\bea d\left( \ta - f\right) &=& 0 \, \rightarrow \, \ta_{\mu} =
f_{\mu} + \p_{\mu} \tilde{\phi} \label{eq1}\\
d\left( \tb - g\right) &=& 0 \, \rightarrow \, \tb_{\mu} = g_{\mu} + \p_{\mu}
\tilde{\psi} \label{eq2}\eea

\bea g_{\mu} &=&  \sqrt{a_2} E_{\mu\alpha}\ta^{\alpha} \label{eq3}\\
f_{\mu} &=& \frac 1{a_0} E_{\mu\nu}\left( a_1 \ta^{\nu} + \sqrt{a_2}\tb^{\nu}\right)
\quad , \label{eq4} \eea

\no with $\tilde{\phi},\tilde{\psi}$ arbitrary functions. The equations
(\ref{eq1}),(\ref{eq2}) and (\ref{eq3}) may be used to eliminate the fields $\ta_{\mu}
, \tb_{\mu} $ and $g_{\mu}$ in terms of $f_{\mu}$ which becomes the only independent
degree of freedom. In this case (\ref{eq4}) becomes the generalized self-dual equation
$f_{\mu} = f_{\mu}^* $ as in (\ref{eqm}). On the other hand, we could have used
(\ref{eq1}) and (\ref{eq2}) to write $f_{\mu}$ and $g_{\mu}$ in terms of $\ta_{\mu}$
and $\tb_{\mu}$ respectively. Accordingly, plugging back the result in (\ref{eq3}) and
(\ref{eq4}) we derive:

\bea \tb_{\mu} &=& \sqrt{a_2} E_{\mu\nu}\ta^{\nu} +
\p_{\mu}\tilde{\psi} \label{eq3b} \\
\ta_{\mu} &=& \p_{\mu}\tilde{\phi} + \frac 1{a_0} E_{\mu\nu}\left( a_1 \ta^{\nu} +
\sqrt{a_2}\tb^{\nu} \right) = \p_{\mu}\tilde{\phi}
+ C_{\mu} \label{eq4b}\\
&=& \p_{\mu}\tilde{\phi} + \frac 1{a_0}\left( a_1 E_{\mu\nu} - a_2 \Box\theta_{\mu\nu}
\right) \ta^{\nu} = \p_{\mu}\tilde{\phi} + \ta_{\mu}^* \label{eq4c} \eea

\no Where the combination $C_{\mu}$ is the same one defined in (\ref{c}). Since the
generalized duality transformation is such that $\left( \p_{\mu} \tilde{\phi}
\right)^* = 0 $, it is clear from (\ref{eq4c}) that $\ta_{\mu}^* =
\left(\ta_{\mu}^*\right)^* $ and using $\ta_{\mu}^* = C_{\mu} $ we deduce the
self-dual equation $C_{\mu} = C_{\mu}^* $. Therefore, we can say that the map below
holds at quantum and classical level:

\be f_{\mu} \Leftrightarrow C_{\mu} \label{map} \ee

\no In summary, on one hand we have the equations of motion of the first order version
of the GSD model which can be written as $g_{\mu} = \sqrt{a_2}
E_{\mu\alpha}f^{\alpha}$ and $f_{\mu} = f_{\mu}^*$. On the other hand, the equation
(\ref{eq3b}) teaches us that the combination $\tb_{\mu} $ can be eliminated in terms
of $\ta_{\mu}$ in parallel to the elimination of $g_{\mu}$ as function of the
self-dual field, while the dynamical degree of freedom $\ta_{\mu}$ satisfies
$\ta_{\mu}^* = \left(\ta_{\mu}^*\right)^* $ which is equivalent to $C_{\mu} =
C_{\mu}^* $ and therefore completes the analogy with the GSD model. So in both
theories we have only one independent dynamical vector field which satisfies a
self-duality condition. Thus, we can say that both theories are generalized versions
of the  self-dual model of \cite{TPN}. From the point of view of the non-interacting
MCS fields $A_{\mu}$ and $B_{\mu}$ it is quite surprisingly that there is only one
independent dynamical vector field. The reader may find useful to obtain the equations
(\ref{eq3b}) and (\ref{eq4b}) directly from the 2MCS theory as we do next in order to
clarify this point. Minimizing ${\cal L}_{2MCS}$ we have:

\bea E_{\mu\nu}\left( m_- B^{\nu} - E^{\nu\alpha} B_{\alpha}\right) &=& 0 \,
\rightarrow \, B^{\nu} = \frac{E^{\nu\alpha}B_{\alpha}}{m_-} + \p^{\nu} \psi
\label{eq5} \\
E_{\mu\nu}\left( m_+ A^{\nu} + E^{\nu\alpha} A_{\alpha}\right) &=& 0 \, \rightarrow \,
A^{\nu} = -\frac{E^{\nu\alpha}A_{\alpha}}{m_+} + \p^{\nu} \phi \label{eq6} \eea

\no The general solutions (\ref{eq5}) and (\ref{eq6}) lead to $- m_+^{3/2} A^{\nu} -
m_-^{3/2} B^{\nu} = E^{\nu\alpha}\left( \sqrt{m_+} A_{\alpha} - \sqrt{m_-} B_{\alpha}
\right) - \p^{\nu}\left(m_+^{3/2}\phi + m_-^{3/2}\psi \right)$ which is equivalent to
equation (\ref{eq3b}), i.e., $\tb^{\nu} = \sqrt{a_2}E^{\nu\alpha}\ta_{\alpha} +
\p^{\nu}\tilde{\psi} $ . This confirms that we can treat $E^{\nu\alpha}\ta_{\alpha}$
as the only independent dynamical vector field in the 2MCS model. Analogously, from
(\ref{eq5}) and (\ref{eq6}) we have $\sqrt{m_+}A_{\mu} - \sqrt{m_-}B_{\mu} = -
E_{\mu\alpha}\left(A^{\alpha}/\sqrt{m_+} + B^{\alpha}/\sqrt{m_-}\right) $ $+ \p_{\mu}
(\sqrt{m_+}\phi - \sqrt{m_-}\psi)$ from which we can derive $ \ta^{\mu} =
E^{\mu\alpha}\left\lbrack \ta_{\alpha} (m_+ - m_-) +
\tb_{\alpha}/\sqrt{a_2}\right\rbrack /(m_+m_-) + \p_{\mu}\tilde{\phi}$ which is
equivalent to equation (\ref{eq4b}) and consequently we deduce the generalized
self-duality equation $C_{\mu}=C_{\mu}^*$ with $C_{\mu}=A_{\mu}^*$. The quantities
$\tilde{\psi},\tilde{\phi}$ are  of course linear combination of $\psi$ and $\phi$.

At last, we briefly comment on the coupling of the GSD model to matter and its dual
gauge theory. We notice that the GSD model is not a gauge theory, so there is no
reason to minimally couple it to $U(1)$ matter. In particular, it is natural, see
comments in \cite{jpa}, to consider a linear coupling of the self-dual field to a
$U(1)$ matter current which may represent fermions or bosons (scalars). By repeating
the steps which have taken us from the GSD to the 2MCS model and substitute $j^{\mu}$
by $j^{\mu}_{\rm matter}\, $ it is easy to verify that we have the following duality
relation when we include matter:

\bea {\cal L}_{GSD}(f) \, + \, {\cal L}_{\rm matter} \, + \,f_{\mu} j^{\mu}_{{\rm
matter}} \, &\Leftrightarrow & \, {\cal L}_{2MCS}(A,B) \, + \, {\cal L}_{{\rm matter}}
\, - \,
\frac{j^{\mu}_{{\rm matter}}j_{\mu\,{\rm matter}}}{4 a_0}\nn \\
&+& \frac{j^{\mu}_{{\rm matter}}\emn\p^{\nu}}{\sqrt{a_2(m_+ +
m_-)}}\left(\frac{A^{\gamma}}{\sqrt{m_+}}   + \frac{B^{\gamma}}{\sqrt{m_-}}\right)
\label{matter} \eea

\no Therefore, the dual gauge theory contains a Thirring-like term plus a non-minimal
coupling of the Pauly-type as in \cite{GMS,Bazeiajpa}. Only a specific gauge invariant
linear combination of the Chern-Simons fields couples to the matter current. We
interpret the appearance of non-renormalizable interactions in the dual gauge theory
as a consequence of the bad ultraviolet behavior of the self-dual propagator, which
becomes a constant for large momenta in spite of the presence of the Maxwell term.

\section{Conclusion}

We have suggested here in a systematic way a new master action which correctly
interpolates between a generalized self-dual model (GSD) and its dual gauge theory
consisting of a couple of non-interacting Maxwell-Chern-Simons  fields of opposite
helicities (2MCS) which is local and ghost free as opposed to previous proposals. The
master action suggested here, by construction, assures that the dual theories have the
same spectrum which is not a general feature of the master action approach as
explained in \cite{jhep}. Another key ingredient was the reduction of the second order
Maxwell term to first order by means of an auxiliary vector field $g_{\mu}$ besides
the dynamical self-dual field $f_{\mu}$. This approach allowed a natural parallel with
the two fields of the 2MCS theories thus, explaining the apparent doubling of fields
on one side of the duality. It turns out that both GSD and 2MCS models have a
superfluous vector field which can be eliminated in favor of a gauge invariant
dynamical vector field whose equation of motion can be written as a generalized
self-duality condition.

Furthermore, we have found a  map, see (\ref{map}), between the dual theories which
holds at classical and quantum level. In the opposite direction one can also calculate
the relevant gauge invariant correlators of the 2MCS theory from the GSD model plus
contact terms. Our work demonstrates a complete equivalence between those models and,
differently from \cite{BK2}, no explicit gauge condition has been fixed.

It is possible (under investigation now) that other soldered theories, see
\cite{cwreview} for more examples, can be similarly ``unsoldered'' as we have done
here. In particular, it is tempting to investigate by an interpolating master action
the doubling of fields in the electric-magnetic duality invariant Schwarz-Sen model
\cite{SS} in $3+1$ dimensions.

\section{Acknowledgements}

This work was partially supported by \textbf{CNPq}.  We thank Alvaro de Souza Dutra
for useful discussions and bringing the soldering literature to my knowledge.

\end{document}